\documentclass{webofc}

\usepackage[varg]{txfonts}   
\usepackage{hyperref}
\usepackage{url}
\hypersetup{colorlinks=true,citecolor=blue,urlcolor=blue,linkcolor=blue}
\begin{document}

\title{Experimental overview: quarkonia as probes for deconfinement}
%
%

%
\author{\firstname{Michael} \lastname{Winn}\inst{1,3}\fnsep\thanks{\email{michael.winn@cea.fr}} 
}
\institute{DPhN/Irfu, CEA, Université Paris-Saclay
          }

\abstract{The current status of the experimental investigation of quarkonia in heavy-ion collisions as probes for deconfinement is reviewed. The emphasis is put on the qualitative understanding of charmonia ($c\bar{c}$ bound states) and bottomonia ($b\bar{b}$) based on the results in the past few years. In particular, the observation of non-primordial J/$\psi$ ($c\bar{c}$) production as signature of deconfinement based on different observables  and the sequential suppression patterns of the $\Upsilon$ ($b\bar{b}$) states are discussed. An outlook focusing on the time scale until the next "Quark Matter" conference is provided. 
}
\maketitle
\vspace{-0.5cm}

\textbf{Introduction}

\label{intro}
Quarkonium states, bound states of heavy quark-antiquark pairs, are considered as a key observable for the investigation of the quark-gluon plasma (QGP) in heavy-ion collisions, in particular for the observation of deconfinement itself. These states can be seen as the 'hydrogen atom' of quantum chromodynamics (QCD):  the heavy quarks  represent the best approximation in 
nature of static color charges. In a pure  gluon theory without light quarks, the free energy of a heavy quark and a heavy antiquark can be directly related to the order parameter for the deconfinement transition, the Polyakov loop~\cite{Rischke:2003mt}. However, lattice QCD indicates that the transition between hadronic matter and the QGP is a rapid cross over for vanishing baryochemical potential~\cite{Borsanyi:2013bia,HotQCD:2014kol}, 
the part of the phase diagram probed  in heavy-ion collisions at colliders and in the early universe according to standard cosmology~\cite{Braun-Munzinger:2008szb}. Therefore, the characterization of the deconfinement transition from the QGP towards hadrons in nucleus-nucleus collisions is a central quest for heavy-ion physics.
The modification of quarkonia production with respect to the expectation in absence of a deconfined medium is  hence considered as a key signature.

Heavy quarkonium represents at hadron colliders only a small fraction of the total heavy-quark pair production, about 1\% of the total $c\bar{c}$ production hadronizes into charmonium~\cite{Schweda:2014tya}. 
Whereas the production of quarkonia itself is sensitive to the strongly interacting matter produced in the collision, the number of heavy quarks and heavy antiquarks is approximately conserved after the initial stages of the collision throughout the lifetime of the produced medium. 
Hence, the normalization of quarkonium production to the production of the sum of all open heavy-flavor states is the most model-independent observable to quantify enhancement or suppression of quarkonium production with respect to the hadronization without the presence of strongly interacting matter. We will focus in these proceedings on transverse momentum integrated and low-transverse momentum observables ($M_{quarkonium}$ not small compared to $p_{T,quarkonium}$). In this kinematic domain, the heavy-quark system has only a small boost with respect to the local fluid rest frame. Partial or full thermalization scenarios are typically applied in model calculations. Measurements at large transverse momentum, where the energy loss of a parton (mostly gluons) or the heavy-quark-anti-quark pair can be assumed to play a crucial role, are only touched upon.

Until this day, measurements in proton-proton and in proton-nucleus collisions have been important for QGP-quarkonia physics: they did not only serve as  reference measurements for nucleus-nucleus collisions, but they have provided new observables, puzzles and new ideas for the  nucleus-nucleus collision program. Although the focus of these proceedings is on nucleus-nucleus collisions, proton-proton and proton-nucleus collision results are discussed in this perspective.
In nucleus-nucleus collisions, the available measurements are restricted to the dilepton decay channels, a limitation arising from the large combinatorial background for multi-body decays involving hadrons and/or photons, whereas in proton-proton and in proton-nucleus collisions hadronic as well as dilepton$+$photon decay channels have been exploited, see e.g. the measurement of $\eta_c$~\cite{LHCb:2014oii,LHCb:2019zaj,LHCb:2024ydi} and of $\chi_c$~\cite{CDF:1997uzj,LHCb:2012af,CMS:2012qwg,PHENIX:2013pmn,LHCb:2021zwp,LHCb:2023apa,CMS:2025rnw}, a discussion of the feed-down chains is provided in~\cite{Lansberg:2019adr}. After the discussion of the currently available results, we will provide a wishlist of opportunities beyond these by now quite mature measurements at colliders. 

\vspace{0.3cm}

\textbf{Charmonium regeneration: signature of deconfinement}
\label{sec-1}

In the charmonium sector, we can access experimentally in nucleus-nucleus collisions with good precision the lowest lying vector state, the J/$\psi$. Measurements of the second vector state $\psi$(2S) below the open charm threshold are limited at colliders in nucleus-nucleus collisions. Only very recently, first measurements integrating over the full transverse momentum range are available.    

Prior to data from RHIC and the LHC, the non-primordial production of J/$\psi$ within or from the deconfined medium has been advertised as a signature of deconfinement~\cite{Braun-Munzinger:2000csl,Thews:2000rj} 
at collider energies: a large initial charm quark density and charm quark conservation after production yields to a larger production rate of heavy quarkonium than expected in suppression scenarios. Two different scenarios of non-primordial production can be distinguished: the statistical hadronization model (SHMc) assuming the complete production at the phase boundary first introduced in~\cite{Braun-Munzinger:2000csl}, and transport models first introduced in~\cite{Thews:2000rj} assuming the production and destruction of bound states during the deconfined  phase. 

\begin{figure}
    \centering
    \includegraphics[width=0.3\linewidth]{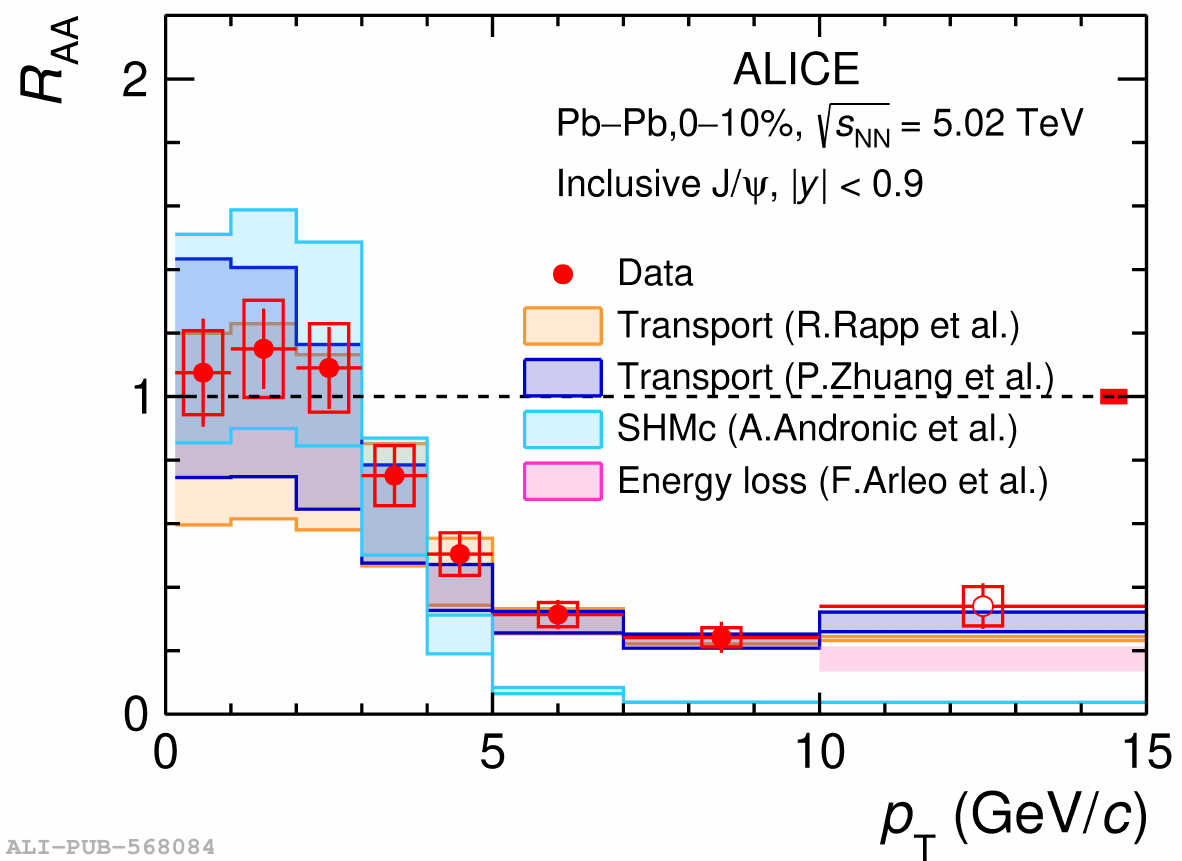}
\includegraphics[width=0.33\linewidth]{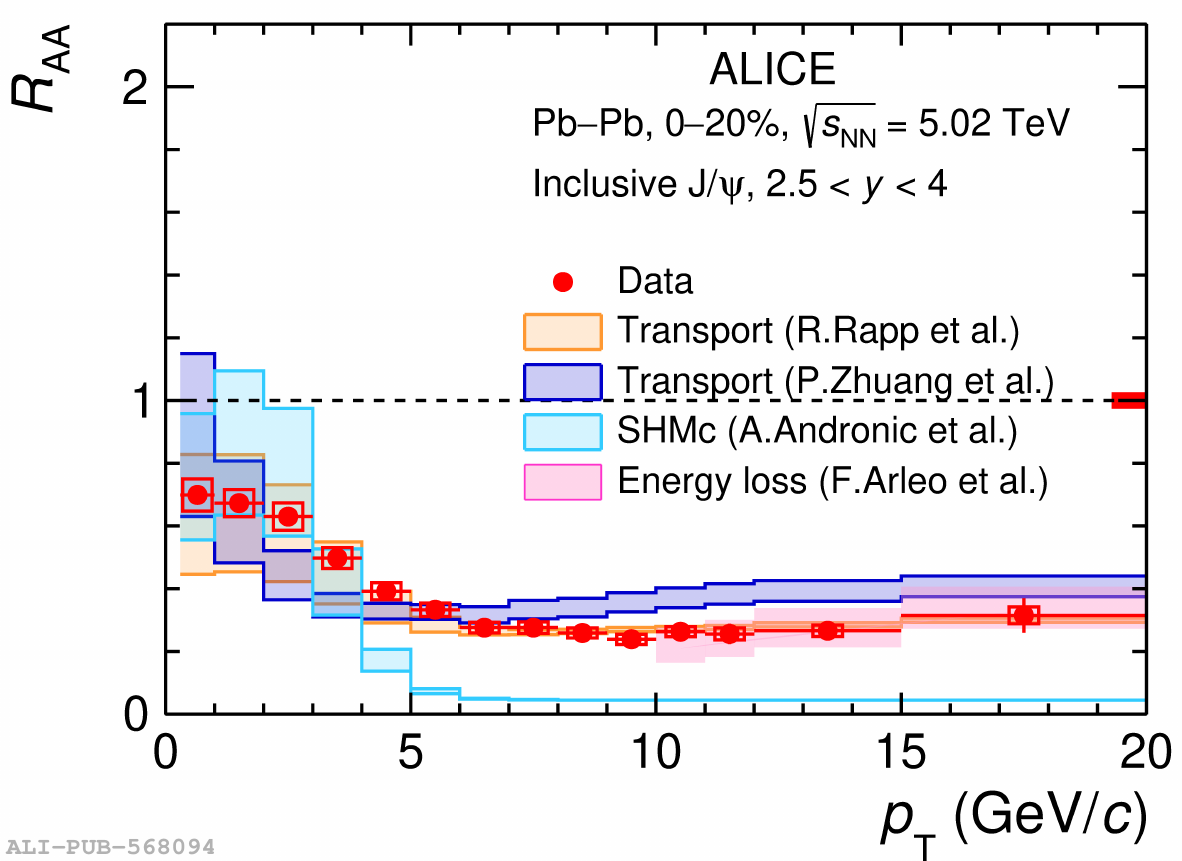}
\includegraphics[width=0.3\linewidth]{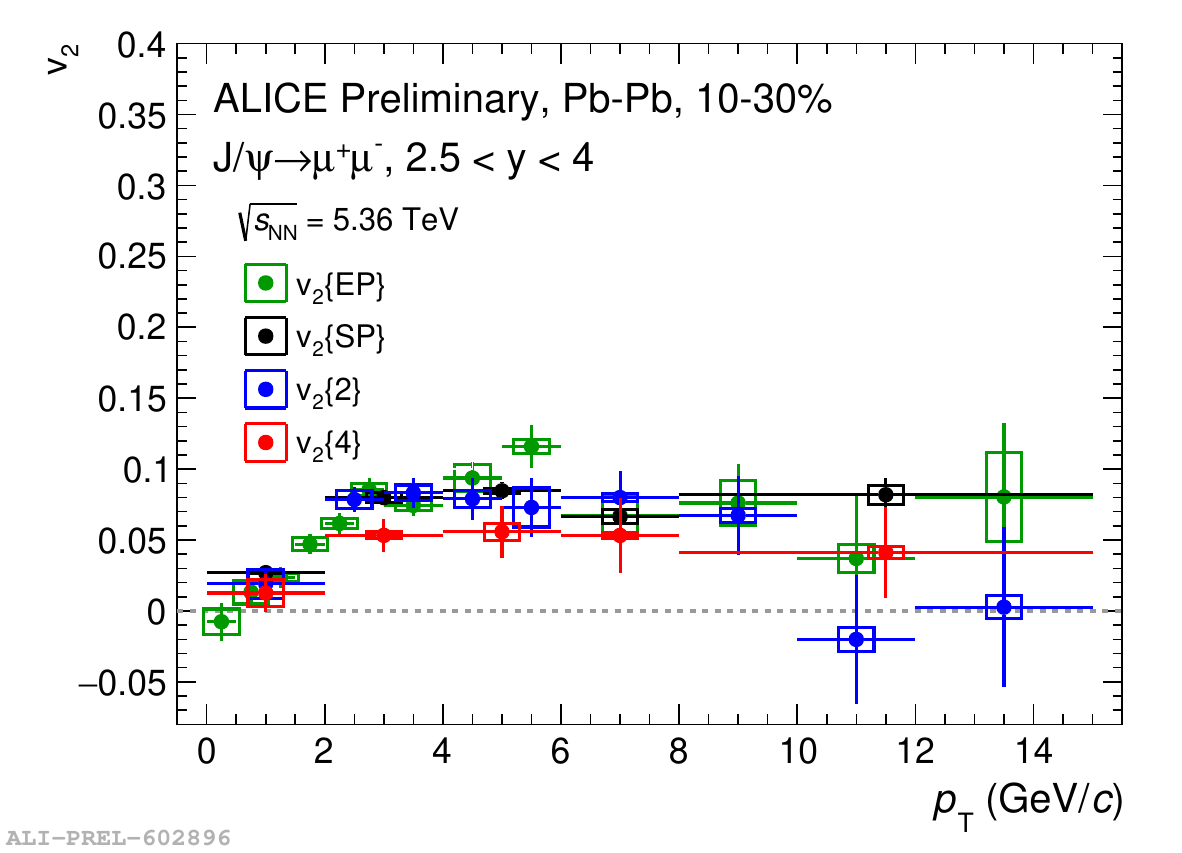}
    \caption{Results on the inclusive J/$\psi$ nuclear modification factor from ALICE at midrapidity (left) and at forward rapidity (middle) taken from \cite{ALICE:2023gco} compared to model calculations~\cite{Andronic:2019wva,Zhao:2007hh,Zhou:2014kka}, 
     see also in Ref.\cite{ALICE:2023hou} for the separation of the prompt and the non-prompt part. Preliminary results on the elliptic flow of J/$\psi$ from Run 3 are shown on the right. }
    \label{fig:legacyRun2jpsi}
\end{figure}

The LHC Run 2 results confirm the regeneration scenario with a variety of observations: the larger nuclear modification factor at midrapidity than at forward rapidity for transverse momentum integrated nuclear modification factors (related to the larger charm quark density at midrapidity), an increase of the nuclear modification factor towards the lowest transverse momenta (indicating the 'thermal' origin of the quarkonia states), a sizeable anisotropic flow of J/$\psi$, the centrality dependence and the collision energy dependence. We show in Fig.~\ref{fig:legacyRun2jpsi} the result on the nuclear modification factor from Run 2~\cite{ALICE:2023gco} and the preliminary results for the azimuthal anisotropy from Run 3 that are compatible with the results  from Run 2~\cite{ALICE:2020pvw}. 

\begin{figure}
    \centering
    \includegraphics[width=0.48\linewidth]{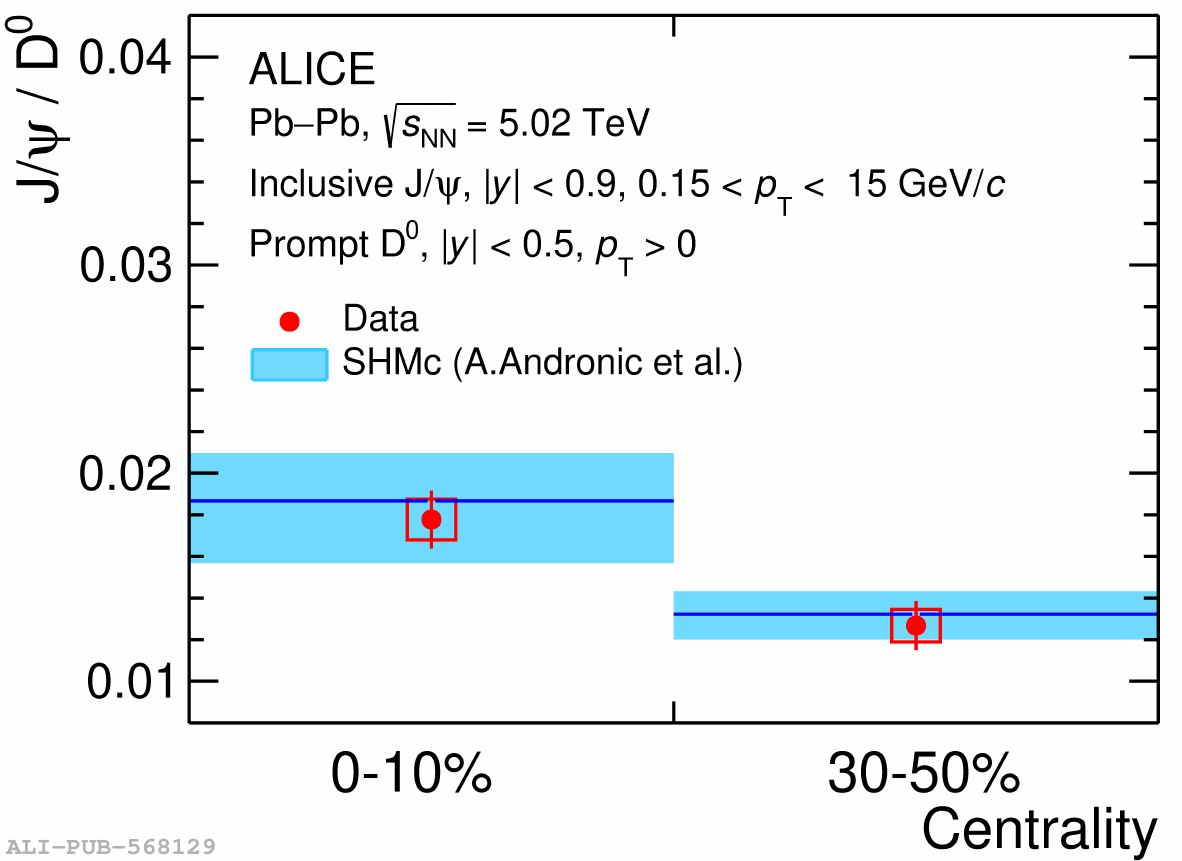}
    \includegraphics[width=0.48\textwidth]{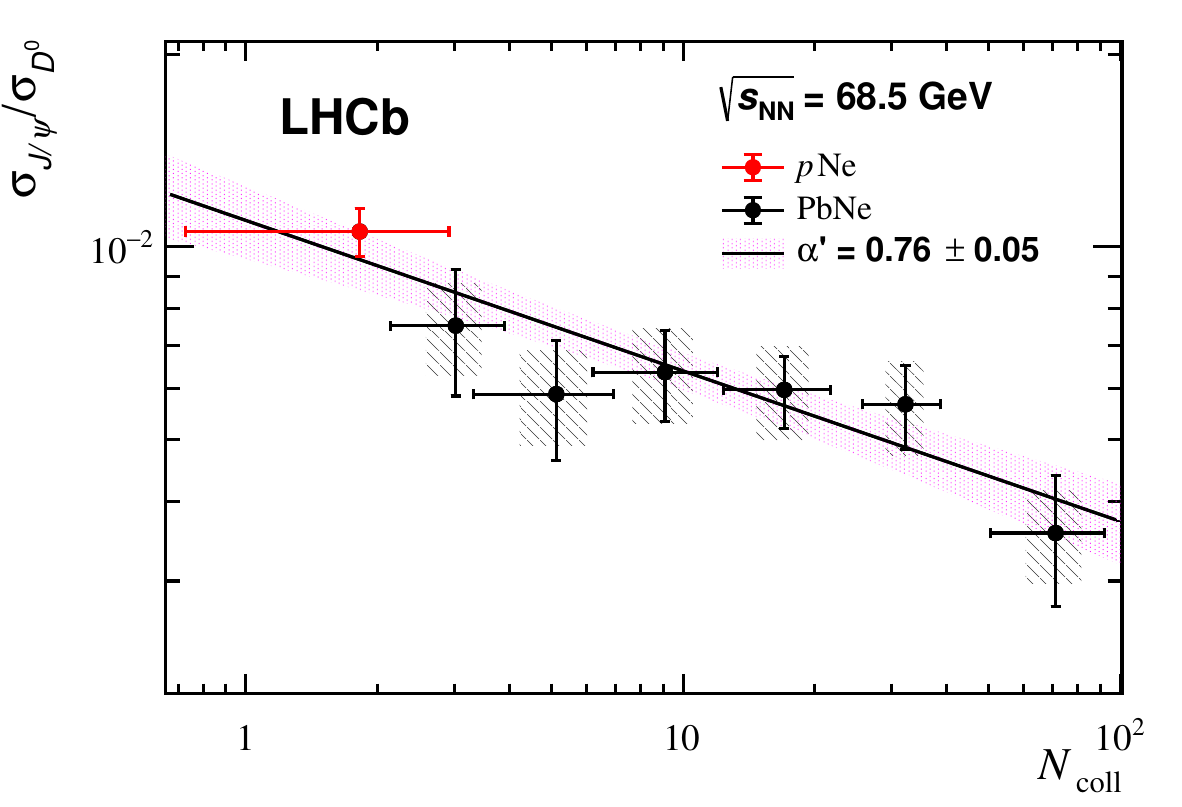}
    \caption{The production ratio of J/$\psi$ over $D^0$ as measured in ALICE at midrapidity compared with the statistical hadronization model~\cite{Andronic:2019wva} taken from Ref.\cite{ALICE:2023gco}. The same ratio as function of $N_{coll}$ in proton-Neon and lead-Neon collisions as measured by LHCb in Ref.~\cite{LHCb:2022qvj}.} 
    \label{fig:DtoJpsiratio}
\end{figure}

However, despite these findings, the observables mentioned above do not allow to discriminate between the two scenarios, production at the phase boundary and continuous destruction and production throughout the QGP lifetime. A key parameter common to both Ansaetze is the total charm production. This observable is not sufficiently well measured to exploit the  prediction difference for the ratio between the J/$\psi$ production and the total charm production. In fact, in order to describe the experimental data at midrapidity, the total charm cross section at midrapidity with the SHMc amounts to $d \sigma_{c\bar{c}} = 0.68 \pm 0.10$ mb, whereas the corresponding value in the  
transport model from R. Rapp and collaborators is as large as $0.93\pm 0.12$ mb\cite{Andronic:2025jbp}. 
Fortunately, the total charm production is a measurable quantity and is a major goal of the ALICE and LHCb measurements in Run 3 on open heavy-flavor including the baryons both in collider and in fixed-target mode. A first step towards this goal is the normalization of the J/$\psi$ to the $D^0$ meson production integrated over transverse momentum. This has been achieved by ALICE in PbPb collisions at $\sqrt{s_{NN}}=$5 TeV in Run 2 and by LHCb in PbNe collisions  in Run 2 $\sqrt{s_{NN}}=68.5$ GeV (Fig.~\ref{fig:DtoJpsiratio}). The ratio observed by ALICE is in line with the prediction by the SHMc.  
It is noteworthy that the confrontation of the LHCb fixed-target data with SHMc and with transport models should be pursued urgently. 

Alternatively to the normalization to the total charm production in nucleus-nucleus collisions, it is possible to constrain the total charm production indirectly by the means of proton-nucleus or $\gamma$-nucleus collisions 
under the assumption that the production of charm is dominated by the production in the initial hard scatterings and that the modification of the parton distribution functions (PDFs) between a nucleon and a nucleus is the dominant effect on the total charm production in nucleus-nucleus collisions compared to proton-proton collisions and that this modification can be factorized in a modification per nucleus, see in ref.~\cite{Apolinario:2022vzg} for a discussion of these assumptions and possible confounding effects. 
In this scenario, we can use $p/\gamma$-nucleus data to constrain nuclear PDFs (nPDFs). They can then be used to calculate the total charm production in nucleus-nucleus collisions based on the total charm measurements in proton-proton collisions and the nPDFs to calculate the total heavy quark production in nucleus-nucleus collisions. In this context,   a strong suppression of inclusive D-meson production in proton-lead collisions is observed at forward rapidity, e.g. this data-set used in different nuclear PDF fits~\cite{LHCb:2017yua}~\footnote{The data at midrapidity is consistent with unity, however exhibits about 3 times larger uncertainties~\cite{ALICE:2019fhe}.} with respect to the naive hard scattering scaling from pp collisions. It is remarkable that the suppression of the photoproduction of coherent J/$\psi$ with respect to the impulse approximation  at large $\gamma-$nucleon energy in  $\gamma$-lead collisions~\cite{ALICE:2023jgu,CMS:2023snh}~\footnote{The recently released ATLAS preprint~\cite{ATLAS:2025aav} with the main results  being presented at this conference casts a doubt on this universal picture and calls for further experimental clarification with respect to the findings of ALICE and CMS.}, a collision type with other complications in view of nPDF modification interpretation. 
Both hadro- and photoproduction in general~\footnote{including inclusive photoproduction being recently persued as well in UPC, see the CMS results~\cite{CMS:2024ayq} presented at this "Quark Matter" conference, now available as preprint~\cite{CMS:2025jjx} } are subject to large factorization scale uncertainties. Progress on the experimental side on these subjects can be anticipated by new measurements including beauty production with better statistical precision than currently available, e.g.~\cite{LHCb:2017ygo,LHCb:2019avm}, with future higher statistics pPb runs 
 being less sensitive to these uncertainties. 
Finally, the already available pPb collision data sets enabled the study the feed-down chains of charmonium towards the J/$\psi$ in detail by measurements of prompt $\chi_c$ state and of prompt $\psi$(2S). The LHCb collaboration shows that the $\chi_c$ feeddown to J/$\psi$ in pPb at forward rapidity~\cite{LHCb:2023apa} 
(charmonium detected at positive rapidity defined by the flight direction of the incoming proton) is compatible with the one observed in proton-proton collisions, whereas at backward rapidity, a larger $\chi_c$ contribution to J/$\psi$ is found. The latter finding is explained by the $\psi$(2S) measurement~\cite{LHCb:2024taa}, 
where a larger suppression is observed at backward rapidity than at forward rapidity.  The suppression of $\psi$(2S) is modeled with rescattering of the $\psi$(2S) with comoving particles~\cite{Ferreiro:2014bia} 
or gluons~\cite{Ma:2017rsu}. 

Given the observations in the correlation and hadronization sector~\cite{Grosse-Oetringhaus:2024bwr} 
indicating a continuous emergence of heavy-ion signature observations in proton-proton and proton-lead collisions with increasing number of produced particles, quarkonium   production ratios have been investigated in detail as function of multiplicity of produced final state particles in proton-proton and proton-nucleus collisions. At this "Quark Matter" conference, a new preliminary result by CMS~\cite{CMS:2025rnw} indicates no multiplicity dependence of high-$p_T$ $\chi_c \to J/\psi + \gamma$  over $J/\psi$ production above $p_{T,jpsi}=6.5$~GeV/$c$. However, for $\psi$(2S) production CMS and LHCb show a significant reduction of the $\psi$(2S) over J$\psi$ ratio in pPb collision within increasing multiplicity both at high-$p_T$ (CMS)~\cite{CMS:2025oqe} and $p_T$-integrated (LHCb)~\cite{LHCb:2025elk} as well as in pp collisions~\cite{LHCb:2023xie}\footnote{An earlier measurement from ALICE did not show a significant effect neither in pp nor in pPb collisions~\cite{ALICE:2022gpu}.}. The result of LHCb in pPb collisions approaches the value observed in nucleus-nucleus collisions. A decrease of the excited $\Upsilon$ states w.r.t. $\Upsilon$(1S) state have been observed as well in proton-proton collisions~\cite{CMS:2020fae,LHCb:2025xrb,ATLAS:2022xar}~\footnote{An ALICE measurement based on lower luminosities is consistent with a flat production ratio as function of charged particle multiplicity~\cite{ALICE:2022yzs}.}.  Qualitatively, these effects can be understood as a consequence of rescattering as explained above. However, a quantitative interpretation and a direct comparison to heavy-ion collisions requires a better comparability between experiment and theory. In particular, the charged particle multiplicity should be provided by an absolute correction on particle level on the experimental side and a full event modeling on the theory side to take into account the different acceptances. Ideally, a normalization of the charmonium state production to the total charm production could be also envisaged. 

In addition to the normalization to total charm production, better discriminative power between the two scenarios for charmonium production in heavy-ion collisions can be provided by the measurement of the excited charmonium states in nucleus-nucleus collisions.
First measurements in central collisions have been published recently by ALICE~\cite{ALICE:2022jeh} and first publication of the $p_T$ integrated yield of prompt $\psi$(2S) over J/$\psi$ production from LHCb~\cite{LHCb:2024hrk}
The current precision of the published data does not allow definite conclusions. A new preliminary result from ALICE from Run 3 is compatible with the previous result, but provides a better statistical precision. The community is looking forward to the final result and also to an update from LHCb based on Run 3. In this context, it is also worth noting that the $\psi$(2S) has been measured at RHIC in central nucleus-nucleus collisions for the first time by STAR shown as a preliminary in the last "Hard Probes" conference and recently available as preprint~\cite{STAR:2025imj}, which is shows evidence for a stronger nuclear suppression of $\psi$(2S) than for J/$\psi$ compatible with the result by ALICE. 

At large transverse momentum, charmonium states should be considered as part of a jet as other high-$p_T$ hadrons. In this kinematic domain, they are typically produced from a gluon in the parton shower. They hence represent an interesting opportunity to study phenomena related to jet quenching sensitive to the parton flavor within the parton shower. A first pioneering measurement beyond the usage of charmonium spectra is the measurement of the J/$\psi$ fragmentation function by CMS in PbPb collisions~\cite{CMS:2021puf}. A limiting factor of the usage of charmonium in jets is however the description of its production mechanism in perturbative QCD~\cite{Lansberg:2019adr}. A first measurement of prompt $\psi$(2S) fragmentation functions by LHCb presented at this "Quark Matter" conference~\cite{LHCb:2024ybz} promises to help to clarify the situation further  since the $\psi$(2S) yield do not contain feed-down contributions from different quarkonium states and is hence easier to interpret than the J/$\psi$ fragmentation function.

\vspace{0.3cm}

\textbf{Bottomonium suppression: signature of deconfinement}

In addition to charmonium, bottomonium production measurements became available at the start of LHC both from LHC and RHIC. Due to its higher mass, kinetic thermalization is expected to be more difficult to achieve. On the theory side, the larger mass allows a better separation of scales between system-intrinsic scales (mass, quark momentum, quark kinetic energy) and the system scales (temperature, expansion).  The modern theoretical description may be summarizes with the picture that the QGP acts as a sieve for the traversing quarkonium state~\cite{Rothkopf:2019ipj}. Generically, this medium-state interaction is assumed to be responsible for a sequential suppression of the various states:  a significantly-reduced production for less tightly bound states than for the deeply bound states. 

\begin{figure}
    \centering
    \includegraphics[width=1.0\linewidth]{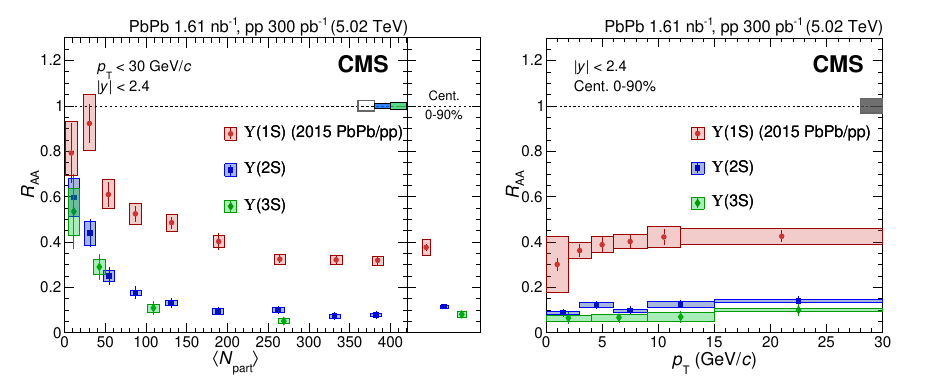}
    \caption{Suppression hierarchy between the three vector states below open beauty threshold as measured by CMS~\cite{CMS:2023lfu}. } 
    \label{fig:Upsilon}
\end{figure}

On the experimental side, the results from CMS~\cite{CMS:2023lfu}, ATLAS~\cite{ATLAS:2022exb} and ALICE~\cite{ALICE:2018wzm} indicated a sequential suppression between the $\Upsilon$(1S) and the $\Upsilon$(2S). CMS measured recently for the first time the $\Upsilon$(3S)~\cite{CMS:2023lfu} that is even more strongly suppressed, see Fig.~\ref{fig:Upsilon}. In addition to the production yield, first elliptic flow measurements  with large uncertainties from Run 2 have been published~\cite{ALICE:2019pox,CMS:2020efs} indicating a $v_2$ compatible with zero at the lowest transverse momenta in line with expectations even of fully thermalized beauty since the velocity field of the fluid only induces a  rise of $v_2$ at relatively large transverse momenta for this heavy particle. ALICE showed at this "Quark Matter" conference a first preliminary measurement based on Run 3 confirming the earlier findings. 

\vspace{0.3cm}

\textbf{LHC Run 3 collider and fixed-target collisions, sPhenix at RHIC: a wishlist for next "Quark Matter" conference}

The large delivered luminosities for all experiments, the strongly increased capability of LHCb to measure up to large number of particles, the LHCb SMOG2 system allowing significantly larger luminosities for the fixed-target program, the luminosity for the central barrel of ALICE by up to 2 orders of magnitude due to the continuous read-out and the addition of the Muon Forward tracker in front of the ALICE muon spectrometer allowing for secondary vertexing promise new high precision data and qualitatively new data with LHC Run 3. In particular, a number of key measurement were so far only feasible in proton-proton or proton-nucleus collisions, whereas they remained inaccessible in nucleus-nucleus collisions until now. 
At this "Quark Matter" conference, we could see first performance figures and first preliminary results from Run 3. In addition, sPHENIX results among them the $\Upsilon$ production at RHIC as a flagship, where so far the precision was limited by statistics and detector performances. 

In this view, a wish-list for next "Quark Matter" conference can be formulated:
\begin{itemize}
\item $p_T$-integrated heavy-flavor normalization for quarkonium in nucleus-nucleus collisions: charm and beauty
\item more constraints from p/$\gamma$-nucleus for nucleus-nucleus collisions in terms of nuclear parton distribution functions
\item excited vector state measurements
prompt $\psi$(2S), $\Upsilon$(3S)
\item first measurements of non-vector states in nucleus-nucleus collisions, probably first the $\chi_c$ state
\item at LHC collider: forward and midrapidity
\item in fixed-target mode LHC and at RHIC.
\end{itemize}
These measurements can be expected both in collider mode at the LHC at forward as well as midrapidity. In addition, the fixed-target mode of LHC and at RHIC will also contribute on this list with the lever arm towards lower collision energies. 
\vspace{0.3cm}

\textbf{Conclusion: Quarkonium - the observable for deconfinement}

With the combination of results from RHIC and the LHC, clear signatures of deconfinement have been ironed out using quarkonium observables: regeneration of charmonium states and the suppression hierarchy observed for the bottomonium vector states. 

We can expect qualitatively new data and improved precision in the near future. We could get already a glimpse with the first $p_T$-integrated results of $\psi(2S)$ and of $\Upsilon$(3S) with LHC Run 2 data. LHC Run 3 (collider and fixed-target mode) and sPHENIX at RHIC results will soon become available. With the measurements of new states, more constraints on the initial state and the total heavy-quark production in nucleus-nucleus collisions, we are in the position to improve our microscopic understanding of quarkonium in the QGP and to constrain non-deconfinement effects. In order to fully profit from this new data and to pave the way for the future programs at the HL-LHC (ALICE3 and LHCb U2, ATLAS and CMS phase-2), a strong dialogue between experiment and theory is required. 


%
%
%
{\tiny 

}

\end{document}